\documentclass[twocolumn,nofootinbib,aps,prl,floats,floatfix,amsmath,amssymb,longbibliography,secnumarabic,
superscriptaddress,preprintnumbers]{revtex4-1}
\usepackage[final]{graphicx}
\usepackage[usenames]{color}
\usepackage{setspace}
\usepackage{amsmath}
\usepackage{amsthm}
\usepackage{mathrsfs}
\usepackage{epsf}
\usepackage{epsfig}
\usepackage{bbm}
\usepackage{bm}
\usepackage{amsfonts}
\usepackage{amssymb}
\usepackage{latexsym}
\usepackage{graphicx}
\usepackage[english]{babel}
\usepackage{multirow}
\usepackage{float}
\usepackage{url}
\usepackage{slashed}
\usepackage{xcolor} 
\usepackage[utf8]{inputenc}
\usepackage{stmaryrd} 
\usepackage{enumitem}
\usepackage{siunitx}
\usepackage{verbatim}
\usepackage[colorlinks,linkcolor=blue,anchorcolor=blue,citecolor=blue,urlcolor=blue,]{hyperref}
\begin{document}

\newcommand{\DC}{{\text{DC}}}
\newcommand{\DF}{{\text{DF}}}
\newcommand{\DR}{{\text{DR}}}
\newcommand{\BH}{{\text{BH}}}
\newcommand{\AC}{{\text{AC}}}
\newcommand{\NS}{{\text{NS}}}
\newcommand{\eff}{{\text{eff}}}
\newcommand{\SM}{{\text{SM}}}
\newcommand{\BSM}{{\text{BSM}}}
\renewcommand{\topfraction}{0.8}

\newcommand{\rv}[1]{\textcolor{black}{#1}}
\newcommand{\rvcom}[1]{\textcolor{red}{#1}}


\title{Imprints of ultralight axions on the gravitational wave and pulsar timing measurement}

\author{Ning Xie}

\author{Fa Peng Huang}
\email{Corresponding Author.  huangfp8@sysu.edu.cn}

\affiliation{MOE Key Laboratory of TianQin Mission, TianQin Research Center for
	Gravitational Physics \& School of Physics and Astronomy, Frontiers
	Science Center for TianQin, Gravitational Wave Research Center of
	CNSA, Sun Yat-sen University (Zhuhai Campus), Zhuhai 519082, China}

\begin{abstract}
The axion or axion-like particle motivated from a natural solution of strong CP problem or string theory is a promising dark matter candidate. We study the new observational effects of ultralight axion-like particles by the space-borne gravitational wave detector and the radio telescope. 
Taking the neutron star-black hole binary as an example, we demonstrate that the gravitational waveform could be obviously modified by the slow depletion of the axion cloud around the black hole formed through the superradiance process. We compare these new effects on the binary with the well-studied effects from dynamical friction with dark matter and dipole radiation in model-independent ways. Finally, we discuss the constraints from LIGO/Virgo and study the detectability of the ultralight axion particles at LISA and TianQin.
\end{abstract}

\maketitle

\textit{\textbf{Introduction.}}---
The ultralight scalar boson, such as the axion~\cite{a1,*a2,*a3,*a4,*a5,*a6,*a7,*Sikivie:2020zpn} or axion-like particle~\cite{Witten2006}, naturally 
correlates particle physics, cosmology, and gravity in many important processes, like the ultralight dark matter (DM)~\cite{CDM,*CDM1982,CDM3} formation and superradiance process~\cite{1971SR,*1972SR,1973SR,Zouros:1979iw,Dolan:2007mj,2020SR} of Kerr black hole (BH). 
The discovery of gravitational waves (GWs) by LIGO/Virgo~\cite{LIGOScientific:2016aoc} has initiated new perspectives to explore these fundamental problems
through various GW experiments including LIGO~\cite{aLIGO2010}, TianQin~\cite{TianQin:2015yph,TianQin:2020hid}, LISA~\cite{LISA:2017pwj}, and Taiji~\cite{Hu:2017mde}.
The well-known superradiance process can extract energy and angular momentum from fast-spinning BH and form the condensed axion cloud~\cite{1971SR,*1972SR,1973SR,Zouros:1979iw,Dolan:2007mj,2020SR} when the axion's Compton wavelength is comparable to the BH size. 
Finally the mass of axion cloud could reach about $ 10 \textrm{\%}$ of the BH mass~\cite{Arvanitaki:2010sy,Arvanitaki:2014wva}. The BH and its axion cloud form a ``gravitational atom''. The axion cloud could lead to various observable effects. One well-studied effect is the monochromatic GW signal from the axion annihilation in the cloud~\cite{Arvanitaki:2010sy,2020SR,Arvanitaki:2009fg,Arvanitaki:2014wva,Arvanitaki:2016qwi,Brito:2017wnc}. LIGO has placed constraints on certain ranges of axion mass~\cite{LIGOScientific:2021jlr,Palomba:2019vxe,Sun:2019mqb,Ng:2020ruv,Yuan:2022bem}, which might be relaxed as discussed in Ref.~\cite{Tong:2022bbl}.

\begin{figure}[h]
	\begin{center}
		\includegraphics[width=0.36\textwidth,clip]{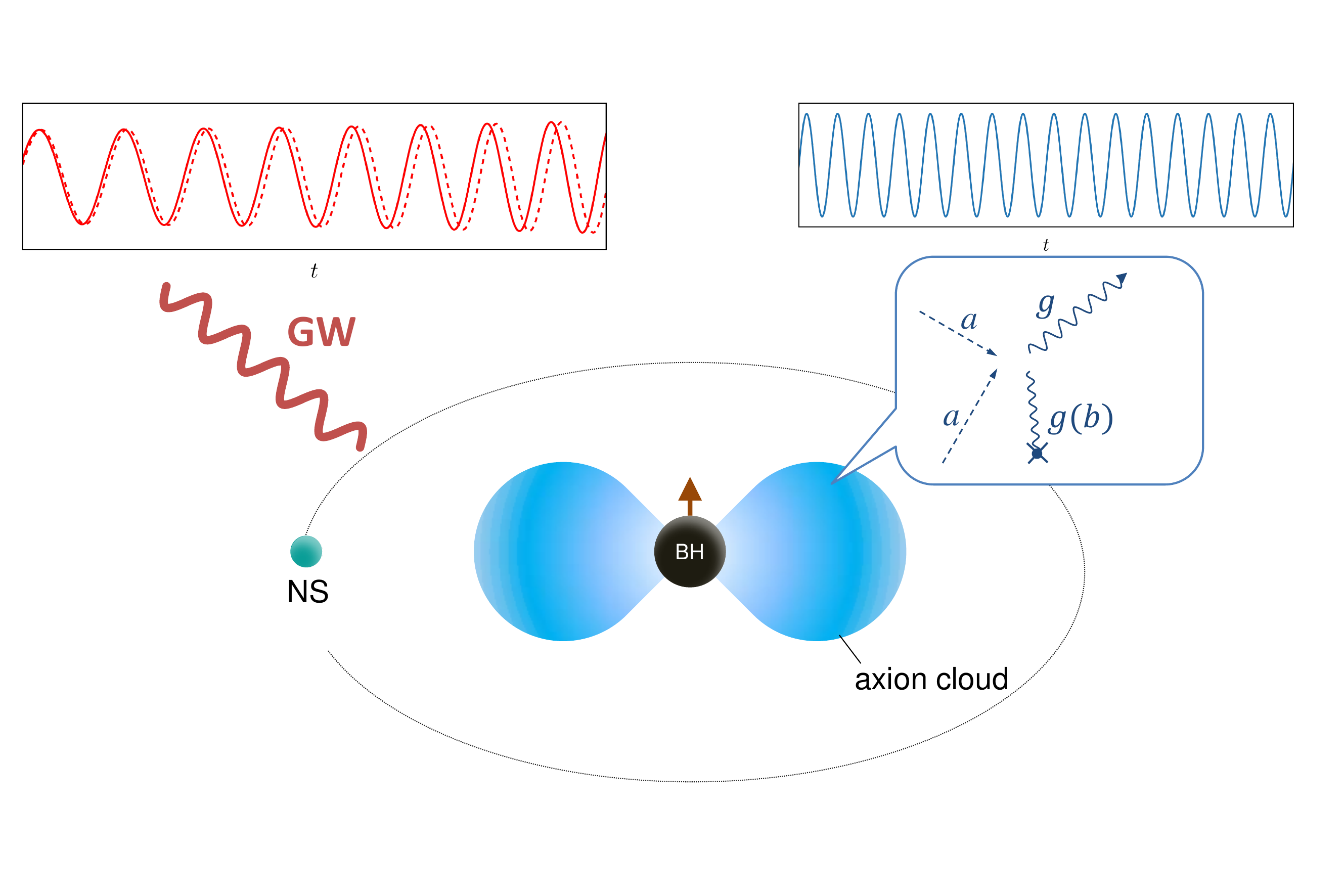}\\
		\caption{{The schematic process with the low frequency GWs generated by the binary system (red) and the high frequency GWs generated in the axion cloud (blue). The depletion of axion cloud can significantly modify the binary GWs.}}
		\label{sc}
	\end{center}
\end{figure}

In this letter, we explore new observational effects to detect the ultralight axion particle by GW and pulsar timing measurement in the neutron star (NS)-BH-like binary {as shown in Fig.~\ref{sc}. Besides the well-studied imprints on the binary GWs from the dynamical friction (DF) and possible dipole radiation (DR),
we notice a missing but the most important mechanism, namely, the mass depletion of axion cloud (DC).
The axions in the cloud induced by the BH superradiant instability could continuously annihilate into gravitons, gradually bringing down the mass of gravitational atom and changing the gravitational field between the binary.  
This new effect could take more energy away} and hence modify the binary evolution in the inspiral stage, which might be observable via GW and pulsar timing measurement.  
Finally we study the distinguishability of the GW waveforms with and without ultralight axions, mostly based on the phase shift of the GW signal at TianQin and LISA. 
In our analysis, we use the units in which $ \hbar= G =c=1 $. 

\textit{\textbf{Binary with ultralight axion.}}---
We start with the BH mass $M_\BH \in  \left[10,~10^3 \right]~{\rm M_\odot} $. 
Such a BH accompanying a NS or another BH could emit low frequency GWs sensitive to space-borne GW detector~\cite{TianQin:2015yph,TianQin:2020hid,LISA:2017pwj,Hu:2017mde}. 
For the binary in the quasicircular approximation without considering the effects from axions, 
the energy loss of inspiral is equal to the power of GW radiation $ \mathcal{P}_{\rm GW} $, 
\begin{align} \label{:energy}
	-\dfrac{\mathrm{d} E_{0}}{\mathrm{d} t} = \mathcal{P}_{\rm GW}  &=  \frac{32 }{5} \mu^2 r^4 \omega^6\,\,, \\
	\mu = \frac{m_\NS M_\BH }{M_\BH+m_\NS}\,\,,  \label{:fre} \ \ \ \omega &= \pi f_{} = \sqrt{\frac{ M_\BH+m_\NS  }{ r^3}}\,\,, 
\end{align}	
with $ \mu $ the reduced mass, 
$\omega$ and $r$ the angular velocity and separating distance of binary, 
$f_{}$ the GW frequency, 
and $m_\NS$ the NS mass. 
With the existence of ultralight axion particles,  more energy could be dissipated through several mechanisms, which will be discussed in the followings.
The benchmark values for $M_\BH$, $m_\NS$, and $f$ are $100~{\rm M_\odot}$, $1.5~{\rm M_\odot}$, and $0.01~\text{Hz}$. 

\textit{\textbf{Depletion of axion cloud.}}---
The boundary condition at horizon of the Kerr BH would lead to imaginary frequency of axion and then the exponential growth of the bound-state axions. In other words, 
the fast-spinning BH spontaneously transforms energy and angular momentum to produce a macroscopic axion cloud and form a gravitational atom through superradiance process~\cite{1971SR,*1972SR,1973SR,Zouros:1979iw,Dolan:2007mj,2020SR}. 
The efficiency of the superradiant instability or the timescale is determined by the 
gravitational fine-structure constant, defined by $ \alpha =  { M_\BH} m_a $ with the axion mass $m_a$.
The superradiance condition~\cite{2020SR} reads $ {\alpha} < \frac{{m\,}\chi}{2 \,( 1+\sqrt{1-\chi^{2}}) } $, with $ \chi $ the dimensionless spin of rotating BH  
and $m$ the magnetic quantum number of axion cloud. 
Assuming $ \alpha = 0.2 $, for $ 2p $ level axion cloud ($\, \left|211\right\rangle$ state), the superradiance rate depends on axion mass and BH spin~\cite{Detweiler:1980uk,Baumann:2019eav}, i.e., 
\begin{equation}\label{rad}
    \Gamma_{\rm SR}=\dfrac{1}{48}g_{p } \left(1-\dfrac{\omega_{2 1 1}}{\Omega_{H}}\right) \chi\hskip 1.0pt{\alpha}^{8}\,{m_a}\,\,.
\end{equation}
Here, $ \omega_{211} $ is the frequency eigenvalues of the axion cloud, $ \Omega_{H} $ is the BH angular velocity at the outer horizon, given by $ \Omega_{H}\equiv \frac{\chi}{2\,M_\BH ( 1+\sqrt{1-\chi^{2}} ) } $, and $ g_{p} = \left(1-\chi^{2}\right)+ [(1-\frac{\omega_{2 1 1}}{\Omega_{H}}) \chi ]^{2} $. 
Besides, the saturation timescale of axion cloud growth is $ \tau_\text{SR} =1~\text{day}\ (\frac{ M_\BH}{100~{\rm M_{\odot}}}) (\frac{{ 0.2}}{_{\,}{\displaystyle \alpha}_{\,}})^9 (\frac{ 0.9}{{ \chi}}) $~\cite{Brito:2017zvb}. 
When exponential growth stops the axion cloud extracts part of angular momentum from BH, and the number of axion particle could be evaluated by~\cite{Arvanitaki:2014wva}
\begin{equation}\label{Naxioncloud}
	N_{\rm{axion}} \sim 10^{78}
	\left(\dfrac{\varDelta \chi}{0.1}\right)\dfrac{1}{m}\left(\dfrac{M_\BH}{100~{\rm M_{\odot}}}\right)^{2}\,\,,
\end{equation}
with $ \varDelta \chi $ the spin difference. 
We consider the annihilation of $ 2p $ level as the main process that radiates graviton, rather than transitions and annihilations of other levels~\cite{Arvanitaki:2016qwi}. The annihilation rate for one pair of axions is approximately as $\Gamma_{a} \simeq \frac{1}{320} \alpha^{12}m_a^3$~\cite{Brito:2014wla} under Schwarzschild approximation, which overestimates the rate. 
As a comparison, the flat-space approximation in Ref.~\cite{Arvanitaki:2010sy} would underestimate this rate. 
In this letter, we adopt numerical results calculated in Kerr metric, where the aforesaid analytical results could be improved~\cite{Brito:2017zvb,Yoshino:2013ofa}. 

The DC effect would gradually reduce the mass of the gravitational atom, and further bring a phase shift to the inspiral GW whose magnitude is determined by the mass depletion rate of gravitational atom's mass $ \tilde{\mathcal{{P}}} $. 
$ \tilde{\mathcal{{P}}} $ depends on the annihilation rate and the number of axions occupying the level, and is up to 
\begin{equation}\label{A}
	\dfrac{\mathcal{\tilde{P}}}{M_\BH}\approx 10^{-6}~{\text{yr}^{-1}} \left(\dfrac{m_a}{10^{-12}~\text{eV}}\right)\,\,  \text{(for }\alpha=0.2 \text{)} \,\, . 
\end{equation}
The duration of the DC effect can be evaluated by $ \tau_\DC \simeq (\Gamma_{a}N_{\rm{axion}})^{-1} \sim 10^{4}~\text{yrs} $~\cite{Arvanitaki:2014wva}. 
Thus the timescale of axion cloud formation is about 6 orders of magnitude shorter than the depletion timescale. 
Therefore for convenience we consider DC process with the assumptions that the axion cloud has grown to its maximum occupation number. 
The DC duration is much longer than the observation time of space-borne GW detector, that helps to ensure a sufficient GW signal duration. 

\textit{\textbf{Dynamical friction and dipole radiation.}}---
The binary system could be surrounded by diffused DM gas composed of ultralight axions or other candidates~\cite{Gondolo:1999ef,Sadeghian:2013laa}. 
The DF of such DM gas with density $ \rho_\text{DM} $ modifies the GW waveform of the binary~\cite{Chandrasekhar:19431,*Chandrasekhar:19432,*Chandrasekhar:19433}. 
To clearly show the effects from DF, we parameterize the DM density around the NS as $ \rho_\text{DM} = 10^n \rho_0  $, 
where $ \rho_0 = 0.3~{\rm GeV/cm^3} $. The factor $n$ depends on the exact location of NS and the DM distribution function, and could be as great as about $ 14 $ for extreme case~\cite{Gondolo:1999ef,Sadeghian:2013laa}.
We assume the BH is approximately static with respect to DM following Refs.~\cite{Chandrasekhar:19431,Eda:2014kra,Zhang:2019eid}. 
The drag force exerted on the NS is opposite to its orbiting velocity $ v$, which is given by 
\begin{equation}\label{DF}
F_\DF=\frac{4 \pi   m_\NS ^2  I(r,v) \rho_\text{DM}}{v^2} \,\,, 
\end{equation}
where $ I(r,v) \sim \mathcal{O}(1) $ is the Coulomb logarithm~\cite{2008C}.
We assume a constant value of $ I(r,v) \approx 3 $ throughout the binary inspiral as in Refs.~\cite{Eda:2014kra,Hannuksela:2019vip}. 

The NS-BH binary could also radiate axions during the evolution at the cost of orbital energy~\cite{Croon:2017zcu,Kopp:2018jom,KumarPoddar:2019jxe,Dror:2019uea}. 
The interacting distance of axion mediated force depends on the angular frequency of axion $ \omega_a $. Namely, the binary radiates axions only when the orbital frequency becomes comparable to, or larger than the axion angular frequency. 
The axion radiation power of this binary is ~\cite{Huang:2018pbu}
\begin{equation}\label{di}
\mathcal{P}_{\text{DR}} \sim  \dfrac{   Q_{}^2 r^2 \omega^4 }{24 \pi } \left( 1-\dfrac{\omega_a^2}{\omega^2} \right) ^\frac{3}{2} \Theta \left( \omega^2-\omega_a^2\right)\,\,, 
\end{equation}
where $ Q_{} $ is the NS's scalar charge and the Heaviside function signifies that the axion radiation turns off when the orbital frequency is much less than the axion angular frequency.  
 The scalar charge can be calculated by $Q = \pm 4 \pi^2 f_a R_{\NS}$ with 
the axion decay constant $ f_a $ and the NS radius $R_{\NS}$~\cite{Huang:2018pbu}.

\begin{figure}[h]
	\begin{center}
		\includegraphics[width=0.30\textwidth,clip]{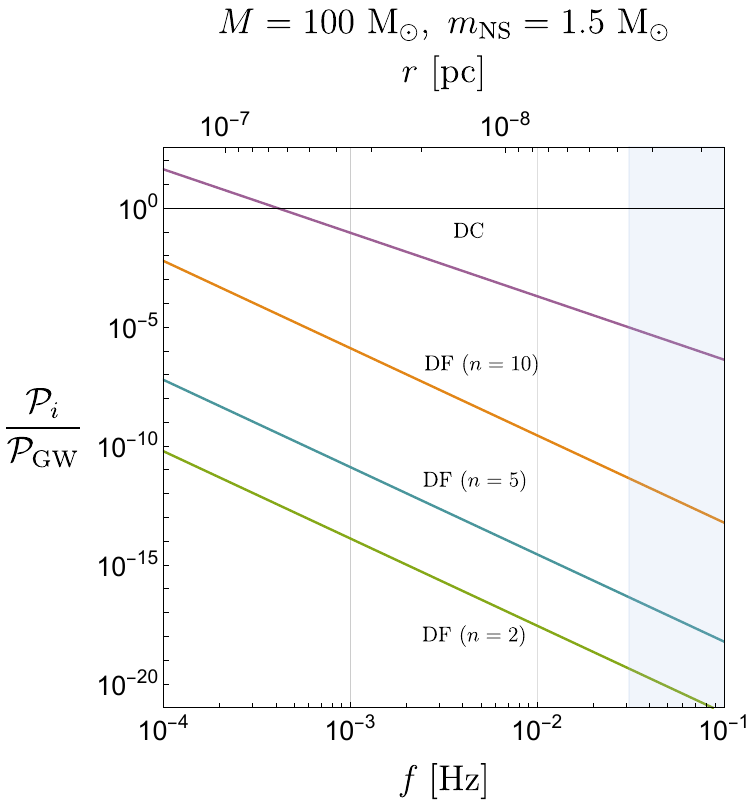}\\
		\caption{The normalized power to GW radiation for different axion effects.  The purple line represents the normalized power of radiated energy from the depletion of axion cloud (DC). The green, blue and orange lines depict the dynamical friction (DF) effect with different DM densities. In low frequency band, the dipole radiation (DR) vanishes. These results are shown with different GW frequencies or orbiting radius. The light blue region means the merger time is less than 5 years. 
		The power of DC effect is comparable to $\mathcal{P}_{\rm GW}$ at low frequency. }
		\label{power}
	\end{center}
\end{figure}

\textit{\textbf{Comparison of three effects.}}---
To extract the axion effects (DC, DF and DR) on NS-BH binary, we investigate and compare their contributions to the binary orbit evolution. 
We focus on the $ 2p $ level axion cloud and omit the influence of other levels. 
We denote $ M $ as the gravitational atom mass. 
Therefore, with the existence of ultralight axions, the DC, DF and DR effects increase the reduction rate of orbital radius, i.e.
\begin{equation}\label{r:evol}
	\frac{\mathrm{d} r}{\mathrm{d} t} = \left(-\dfrac{Mm_\NS}{2 r^2}\right)^{-1}  \left( \mathcal{P}_{\rm GW} + \mathcal{P}_{\DC} + \mathcal{P}_{\DF} + \mathcal{P}_{\DR}  \right) 
\end{equation}
with
\begin{align}
\mathcal{P}_{\text{DC}} &=  \dfrac{ m_\NS}{2 r} \mathcal{\tilde{P}} \,\,, \\
\mathcal{P}_{\text{DF}} &=  \frac{4 \pi   m_\NS^2 I(r,v) \rho_\text{DM}(r)}{\omega r} \,\,.
\end{align}
$ \mathcal{P}_{\text{DR}} $ is given in Eq.~\eqref{di}. 
Notice that for the NS-BH binary, orbital frequency at which the binary begin to generate axion radiation is approximately $ \mathcal{O}(10^2)~\textrm{rad/s} $, so 
when the inspiral binary is less than this frequency the DR effect would not change the GW waveform.

In Fig.~\ref{power}, we show the normalized power of radiated energy for different axion effects. 
The power for each effect is normalized to the GW power $\mathcal{P}_{\rm GW}$ in vacuum and we take binary GW frequency $f $ from $10^{-4}$ to $10^{-1}~\text{Hz}$. 
The purple line represents the normalized power of radiated energy from the DC effect. 
The green, blue and orange lines mark the DF effect for different DM densities with $ n = 2, 5\text{ and } 10 $. 
Since the binary orbital frequency here is much smaller compared to the angular frequency of axion, the DR effect is negligible. 
In the light blue region, the NS-BH binary merger time is less than 5 years, which is the typical operation time of space-borne GW detector. 
We find that the DC effect is at least several orders of magnitude \textit{larger} than the DF effect numerically throughout the whole frequency band we are interested in, even if we assume the DM density with $n=10$. 
Moreover, when the GW frequency is low (e.g., $ f=10^{-3}~\text{Hz} $), 
the power of orbital energy loss induced by the DC effect is comparable with the GW power for the binary in vacuum, and as the binary gets closer, the DC effect becomes relatively weaker. 
In other words, the DC effect could play a \textit{key} role when binary GW frequency is in the low frequency band, while in the high frequency band, the influence of DC effect is small. 

\textit{\textbf{New effects on gravitational wave detector and pulsar timing measurement.}}---
To determine the impact of DC on the GW signals in the space-borne detector, here we calculate the GW phase modification induced by it. 
From Eq.~\eqref{r:evol}, we get 
\begin{equation}\label{f:evol}
\dfrac{\mathrm{d} f}{\mathrm{d} t} = \dfrac{3}{2}\dfrac{\mathcal{\tilde{P}}}{M}f +\dfrac{96 }{5} \pi^{8/3} M_{c}^{5/3} f^{11/3}\,\,, 
\end{equation}
where $ M_{c} $ is chirp mass. 
The time-domain GW phase can be evaluated by $ \phi=\int_0^T 2\pi f(t) \mathrm{d} t  $ with
$ T $ the observation time. 
Based on the result of $\mathcal{\tilde{P}}$, we  obtain
the phase shift in the inspiral waveform for space-borne detector as
\begin{equation}\label{GWphase}
\Delta\phi\sim 15\pi
\left(\dfrac{m_a}{{{10^{-12}}~\text{eV}}}\right)
\left(\dfrac{f_T}{{10^{-2}}~\text{Hz}}\right) 
\left(\dfrac{T}{5 ~\text{yrs}}\right)^2\,\,,
\end{equation}
where $ f_T $ is GW frequency when it was detected. 
We see that, for axion mass about $ {10^{-12}}~\text{eV} $ and GW frequency about $ {10^{-2}}~\text{Hz} $, the phase shift in 5-year observation of space-borne detector~\cite{TianQin:2015yph,LISA:2017pwj} is obviously larger than $2\pi $. 
That indicates it is necessary to consider the DC effect in waveform modeling. 
Note that the metric perturbation from the NS may trigger the reabsorption of the existed axion cloud~\cite{Tong:2022bbl}. 
For the cloud on the $ \left|211\right\rangle $ state, the reabsorption effect happens after the binary leaves the low frequency window. 
Namely, for space-borne detector, it might still be reasonable to model the GW waveform without considering the termination of axion cloud induced by NS. 

The imprints of axions on the NS-BH binary period are also potentially observable through pulsar timing measurement, due to the fact that one of the inspiral objects is a rotating NS. 
Considering the DC effect, for the NS-BH binary we discussed above, at the moment when its orbiting period is $ P=100~\text{s} $, the rate of period change is deviated from the prediction without axion: 
\begin{equation}\label{dpd}
\varDelta\dot{P} = \left| \dot{P} - \dot{P}_\text{vac}\right| \approx 10^{-12}~\text{s/s}\,\,, 
\end{equation}
where $\dot{P}$ and $\dot{P}_\text{vac}$ represent period change rate with and without DC effect, respectively. 
The future measurement precision of $\dot{P}$ is about $10^{-15}~\text{s/s}$ and thus pulsar timing measurement could perhaps detect the axion effects~\cite{Weisberg:2010zz}.

\textit{\textbf{Detectability at TianQin and LISA.}}---
We already point out the GW phase shift caused by the axion effect in low frequency GWs comes up to about $ \mathcal{O}(10)$ (see Eq.~\eqref{GWphase}) and might be detected by space-borne detector. Furthermore, we quantitatively discuss whether TianQin and LISA could detect and unravel this axion effect. 

First, one defines the noise-weighted inner product 
\begin{equation}
\left\langle h_1|h_2 \right\rangle  =2\int_{0}^{\infty}\mathrm{d}f\dfrac{\tilde{h}_1(f)\tilde{h}_2^*(f)+\tilde{h}_1^*(f)\tilde{h}_2(f)}{S_n(f)}\,\,, 
\end{equation}
where $ h_i $ ($ i=1, 2 $) represents time-domain signal, $\tilde{h}_i$ is its Fourier transformation,
and \(S_n(f)\) is the one-sided power spectral density for the instrument noise of the detector~\cite{Finn:1992wt,Cutler:1994ys}. 
The explicit formulas of $S_n(f)$ for TianQin and LISA are given in Ref.~\cite{2018Hu} and \cite{2018Cornish}, respectively. The matched filtering method is used in search of GW signals. 
The signal-to-noise ratio (SNR) is defined as
$ \rho \equiv { \left\langle h|h^{\prime}\right\rangle}/{\sqrt{\left\langle h^{\prime}|h^{\prime}\right\rangle }},  $
where $ h $ is the observed waveform and $ h^{\prime} $ is the template. 
To quantify the difference of two given waveforms, we calculate their match~\cite{Eda:2013gg}, 
\begin{equation}
\mathbb{M}\equiv\frac{\left\langle h|h^{\prime}\right\rangle}{\sqrt{\left\langle h^{\prime}|h^{\prime}\right\rangle\left\langle h|h\right\rangle}}\,\,. 
\end{equation}
With larger modification in realistic waveform, the match $\mathbb{M} $ becomes smaller, and then it is easier to distinguish $h$ from $h^{\prime}$. 
Here, we employ the method in Ref.~\cite{Cutler:2007mi}, and for GW signal with SNR $ \rho $, one takes the following criterion to distinguish two waveforms 
\begin{equation}\label{crit}
\rho >\sqrt{\dfrac{D/2}{1-\mathbb{M}}}\,\,, 
\end{equation}
where $D$ is the number of estimated parameters. 
For TianQin and LISA, typically $D=10$. 
Then we can calculate the critical value of SNR for detecting the phase shift induced by ultralight axion at LISA and TianQin.

Therefore, in order to detect the axion effect, the necessary condition is that the SNR should be larger than the critical value ($ \sqrt{D/[2(1-\mathbb{M})]} $). Besides that, for convenience, we assume that the SNR threshold to claim a detection of GW signal for both TianQin and LISA is about $ 8 $ for a NS-BH-like source. 
Notice that, when the critical value of SNR to detect the axion effect is smaller than the SNR threshold value to detect the GW signals, it can be interpreted the axion effect is significant and the phase shift is observable as long as the GW signals can be detected by the detector. 

\begin{figure}[h]
	\begin{center}
		\includegraphics[width=0.34\textwidth,clip]{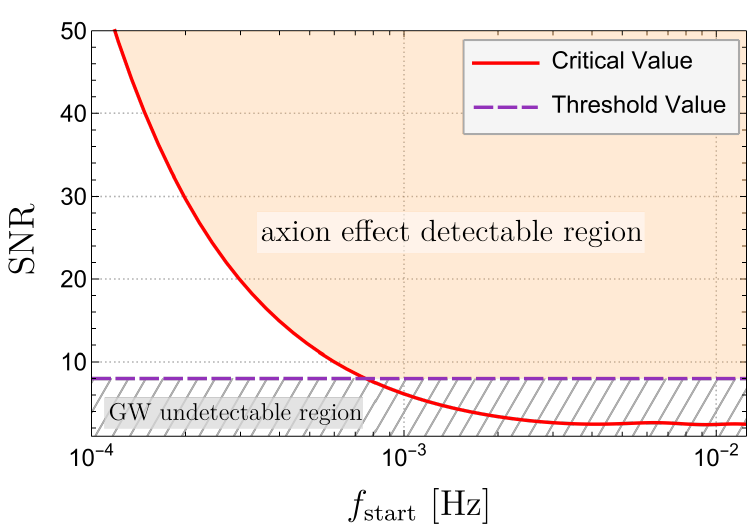}\\
		\caption{The SNR range for detecting the axion effect. The red solid line shows the critical value of SNR to detect the phase modification between the GW waveform with surrounding axions and that in vacuum. The dashed line marks SNR threshold to detect a NS-BH binary at TianQin and LISA. The results are shown with different initial frequency $ f_{\rm start} $. Note that for the NS-BH binary in the light orange region, the axion effect is detectable by GW detector. } 
		\label{requirement}
	\end{center}
\end{figure}

\begin{figure}[h]
	\begin{center}
	\includegraphics[width=0.34\textwidth,clip]{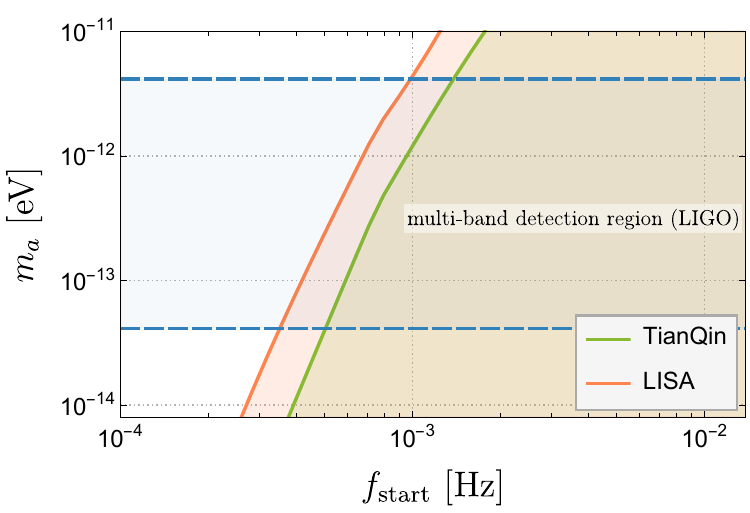}\\
		\caption{The range of source parameters for detecting the axion effect. The green and orange solid lines show the critical parameters at which the phase modification can be detected between the GW waveform with surrounding axions and that in vacuum. The blue dashed line marks threshold to potentially detect high frequency GWs from the axion cloud annihilation at LIGO. The axion effect is detectable by TianQin/LISA for the NS-BH binary in the light green/orange region. We assume the luminosity distance of NS-BH binary is $0.1~{\rm Mpc}$ and the time peroid is 5 years.} 
		\label{region}
	\end{center}
\end{figure}

\begin{figure}[h]
	\begin{center}
		\includegraphics[width=0.34\textwidth,clip]{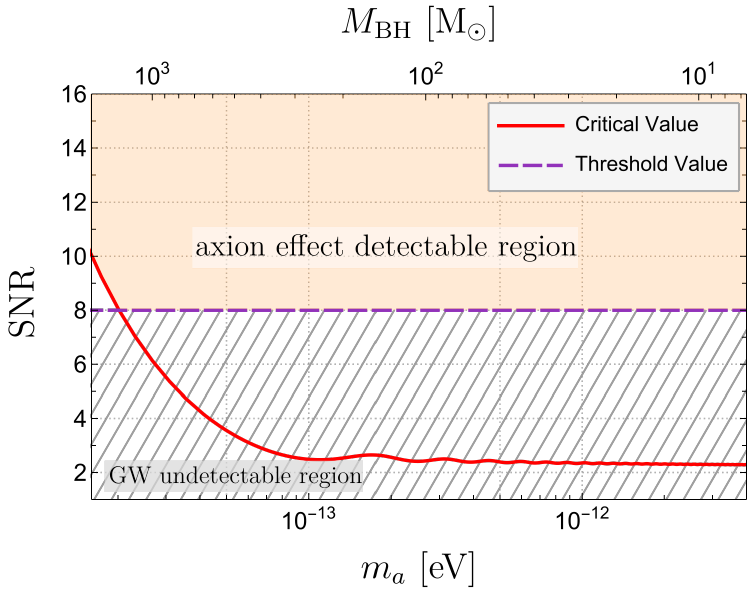}\\
		\caption{The SNR range for detecting the axion effect with different axion masses (bottom-axis) or BH masses (top-axis). The red solid line shows the critical value of SNR to detect the phase modification between the GW waveform with surrounding axions and that in vacuum. The dashed line marks SNR threshold to detect a NS-BH binary at TianQin and LISA. Assume the initial frequency is $  0.01~\text{Hz} $. In the light orange region, the axion effect is detectable by GW detector. } 
		\label{requirement2}
	\end{center}
\end{figure}

\begin{figure}[h]
	\begin{center}
		\includegraphics[width=0.34\textwidth,clip]{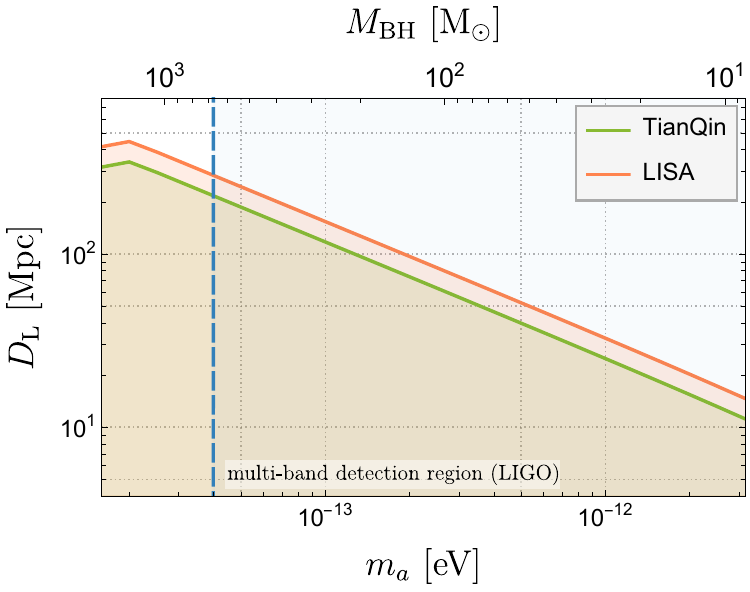}\\
		\caption{The luminosity distance range for detecting the axion effect with different axion masses (bottom-axis) or BH masses (top-axis). The green and orange solid lines show the observational horizon at which the phase modification from axion effects can be detected. The blue dashed line marks lower mass limit to detect high frequency GWs from the axion cloud annihilation at LIGO. 
		Assume the initial frequency is $  0.01~\text{Hz} $. The axion effect is detectable by TianQin/LISA in the light green/orange region. } 
		\label{region2}
	\end{center}
\end{figure}

In Fig.~\ref{requirement}, we show the SNR range in which TianQin and LISA could distinguish the GW waveform with phase shift induced by the DC effect from the regular waveform. 
Here, we assume  $M_\BH=100~{\rm M_\odot}$, $m_\NS=1.5~{\rm M_\odot}$ and $T= 5$ years, and the results are shown with different initial frequency $ f_{\rm start} $. 
The red solid line shows the critical value of SNR to detect the phase modification between the waveform of a NS-BH binary with axion cloud and the putative waveform of the same binary in vacuum. 
The dashed line marks SNR threshold to detect a NS-BH binary source itself at TianQin and LISA. The DC effect is detectable in the light orange region. 
Fig.~\ref{region} shows the range of source parameters in which TianQin and LISA could distinguish the GW waveform with phase shift induced by the DC effect from the regular waveform. 
We adopt the benchmark luminosity distance of NS-BH binary $D_{\rm L}=0.1~{\rm Mpc} $. 
The green solid line shows the critical parameters at which TianQin can detect the phase modification from the surrounding axions and that in vacuum, and the orange solid line shows the result for LISA. 
We consider that TianQin has a half operation period (which is approximately 2.5 year) and LISA has a 3.75-year operation period in 5 years (the assumption is a relatively optimistic expectation~\cite{TianQin:2015yph,LISA:2017pwj}). 
We find that, for space-borne GW detector, the DC effect could be detected from the GW signals more easily when initial GW frequency is larger. 
The critical SNR for discriminating the phase modification is smaller than the SNR threshold to detect GW itself when initial GW frequency is from $ 10^{-3} $ to $ 10^{-2}~\text{Hz} $, which means for this case the phase shift is significant, and it is promising to be discovered as long as the GW is detected.  
That is because the orbital evolution of binary system becomes faster when the frequency is larger. 
Nevertheless, if the initial GW frequency is much larger than $ 10^{-2}~\text{Hz} $, the observation time would be less than 5 years, and consequently the imprint of DC effect would decrease. 
As for considering lower frequency band around $ 2\times 10^{-4}~\text{Hz}  $, the SNR required to distinguish modified waveform is moderate. 
Additionally, in order to show the possibility of multi-band detection, we also represent the region in which LIGO could detect the high frequency GWs from axion cloud in principle. 
In general, the axion effects are potentially observable within the 
axion mass range spanning several orders of magnitude, 
{except for some existing constraints discussed in \textit{Discussion and Conclusion}}.

Furthermore, we consider the SNR required to detect the phase modification with different axion and BH masses in Fig.~\ref{requirement2} . 
Based on the superradiance condition, the cloud with lighter axion can form around heavier Kerr BH. 
We assume the initial frequency is $  0.01~\text{Hz} $ and the integration time is 5 years. 
The red solid line shows the critical SNR to detect the phase modification. 
With more significant DC effect, the critical SNR would be smaller. 
The dashed line represents SNR threshold to detect the NS-BH binary at TianQin and LISA.  The light orange region represents the detectable DC effect.
Fig.~\ref{region2} shows the luminosity distance range in which TianQin and LISA could distinguish the GW waveform with phase shift induced by the DC effect from the regular waveform. 
The green solid line shows the observational horizon at which TianQin can detect the phase modification between the GW waveform with surrounding axions and that in vacuum, and the orange solid line shows the result for LISA. 
We consider the same operation period for TianQin and LISA. 
With more significant DC effect, the critical SNR would be smaller. 
The DC effect is detectable in the light green/orange region for TianQin/LISA. We find that for axion mass ranging from $10^{-13.5} \textrm{-} 10^{-11.5}~\text{eV} $, the phase shift induced from axion effect is detectable, as long as the space-borne detector can detect this binary. 
However, the BH mass determines the strain of GW, so for lower BH mass, the detection horizon becomes closer. 
Also, in order to show the possibility of multi-band detection, we represent the region in which LIGO could also detect the high frequency GWs from axion cloud in principle. 
Generally, the axion effects are potentially observable to the horizon beyond $10~\text{Mpc}$.

\textit{\textbf{{Discussion and Conclusion.}}}---
When the NS and BH  become close, there could be some constraints from ground-based LIGO/Virgo~\cite{Ng:2020ruv}. Nevertheless, the NS could trigger the reabsorption of the axion cloud~\cite{Tong:2022bbl}, and finally the BH could recover its spin. 
In this case the measurement of BH spin could no longer be used to exclude the axion mass range using methods that rely on the existence of axion clouds. 
Therefore, the axion mass constraints in Ref.~\cite{Ng:2020ruv} might be greatly relaxed. 
However, our calculations are consistent with the effect in Ref.~\cite{Tong:2022bbl} as we have discussed in the former part. 
On the other hand, considering the GW directly originating from the annihilation of axion clouds~\cite{Yuan:2022bem}, LIGO data could exclude the mass range  $ \left[1.5,~15 \right]\times10^{-13}~\text{eV} $, $ \left[1.8,~8.1 \right]\times10^{-13}~\text{eV} $ and $ \left[1.3,~17 \right]\times10^{-13}~\text{eV} $, which depend on the BH spin distribution assumption.

We simply estimate the effects of nonlinear self-interaction of axions, such as bosenova~\cite{Yoshino:2012kn,Yoshino:2013ofa}. For the benchmark binary, we find when the axion decay constant $f_a > 5\times10^{16}~\textrm{GeV}$, the bosenova would not occur during the $ 2p $ level axion cloud formation, and the axion radiation effect~\cite{Chavanis:2017loo} in the axion cloud is relatively weak compared to the axion annihilation to gravitons. 
Since ultralight axions around $10^{-13}~$eV discussed in this work mostly favor relatively larger $f_a$, the axion self-interaction effects might be neglected. For small decay constant in some axion models, a detailed study on the effects of nonlinear self-interaction is left for future work.

We have studied the ultralight axions' impact on the GW and the pulsar timing measurement of the NS-BH binary, which are important sources for GW experiments. Besides the extensive studies on the imprints of dynamical friction and dipole radiation, we find that the axion cloud depletion effect dominates over the two effects for the low-frequency GW detection, and LISA or TianQin could be used to explore the ultralight axion particles.
Although we take NS-BH binary as an example, the discussion could be applied to other binaries like binary BHs. This generalization could improve detection probability of axion effect. We expect the future multi-band and multi-messenger observations would help to pin down the properties of ultralight axion.

This work was supported by the National Natural Science Foundation of China (Grant No. 12205387), and the Guangdong Major Project of Basic and Applied Basic Research (Grant No. 2019B030302001).


\end{document}